\documentclass{article}
\begin{document}
\newcommand{\be}{\begin{equation}}
\newcommand{\ee}{\end{equation}}
\newcommand{\bea}{\begin{eqnarray}}
\newcommand{\eea}{\end{eqnarray}}

\title{Space-Time Variation of Physical Constants \\ and the Equivalence Principle}
\author{Kenneth Nordtvedt \\ Northwest Analysis -- 118 Sourdough Ridge Road, Bozeman MT 59715 USA \\ {\it kennordtvedt@imt.net}}
\maketitle

\begin{center}
\section*{Abstract}
\end{center}
\begin{quote}
Location-dependence of physical parameters such as the electromagnetic fine structure constant and Newton's $G$ produce body accelerations which violate universality of free fall rates testable with laboratory and space experiments. Theoretically related cosmological time variation of these same parameters are also constrained by experiments such as lunar laser ranging, and these time variations produce accelerations of bodies relative to a {\it preferred} cosmological inertial frame.   
\end{quote}

Experiments which confirm universality of gravitational free fall rates (UFF) also make impressive statements about the constancy of physical law's key parameters, such as the electromagnetic fine structure 'constant' and Newton's $G$.  If the mass-energy of a body depends on some parameter $\alpha$ which varies in space, for example, there is a force on that body
\be
\vec{F}\;=\;-\;\frac{\partial M(\alpha)}{\partial ln\:\alpha}\;
c^2\;\vec{\nabla}\:ln\:\alpha
\ee
and a corresponding body-dependent acceleration $\delta\vec{a}\;=\;\vec{F}/M$ which will produce violations of UFF. Objects $i$ and $j$ composed of different materials will fall at the differing rates
\be
\frac{|\vec{a}_i-\vec{a}_j|}{|\vec{g}|}\;=\;
\frac{\partial ln(M_i/M_j)}{\partial ln\:\alpha}\;c^2\;\frac{|\vec{\nabla}\:ln\:\alpha|}{g}
\ee 

Based on UFF, Einstein made his generalizing hypothesis --- the Equivalence Principle --- that all physical phenomena in local gravity should be identical as that in an accelerating laboratory.  His prediction followed that two identically constructed clocks located at different altitudes in local gravity must tick at rates $d\tau$ which differ relative to each other as
\be
\frac{d\tau_1}{d\tau_2}\;=\;1\;-\;\frac{\vec{g}\cdot\left(\vec{r}_1-\vec{r}_2\right)}{c^2}
\ee
which relationship integrates in a more global Newtonian gravitational potential to be
\be
\frac{d\tau_1}{d\tau_2}\;=\;1-\frac{U(\vec{r}_1)-U(\vec{r}_2)}{c^2}
\ee
For {\it oscillator} atomic clocks whose frequencies are determined by the difference of two quantum energy levels of a system, $\omega_{kl}\;=\;(e_k-e_l)/\hbar$, a location-dependence of a physical parameter $\alpha$ which participates in the determination of the energy levels will then affect the Equivalence Principle's prediction of clock rate changes.  The clock rate relationship given in Equation (4) is altered to now include an additional clock-dependent factor
\be
\left(\frac{d\tau_1}{d\tau_2}\right)_{kl}\;=\;\left(\frac{d\tau_1}{d\tau_2}\right)_{EP}\;\left(1-
\frac{\partial ln (e_{kl})}{\partial ln\:\alpha}\frac{\partial ln\:\alpha}{\partial U}\right)
\ee
The factor $\partial ln\:\alpha /\partial U$ in Equation (5) and the factor $|\vec{\nabla} ln\:\alpha   |/g$ in Equation (2) are essentially identical, so these two equations can be used to compare the ability of clock experiments and UFF experiments to test for variation of the parameter $\alpha$.  UFF experiments have been superior performers.  Atomic clocks based on different hyperfine transitions will typically have factors $\partial ln(e_{kl})/\partial ln\:\alpha$ differing of order one, while different elements used in UFF experiments will have nuclear electrostatic energy contributions which render the factor $\partial ln\:(M_i /M_j )/\partial ln\:\alpha$ of order $10^{-3}$.  Clock experiments must therefore be able to measure fractional frequency shifts between clocks to precision of a part in $10^{10}$ of the actual gravitational shift in order to compete with UFF experiments now performed to the part in $10^{13}$ precision. But the best clock experiment has reached only the part in $10^4$ precision in testing the EP prediction \cite {ves79}.  Sending atomic clocks on a spacecraft which travels to within a few solar radii of the Sun where the gravitational potential grows to $10^{-6}\:c^2$ could be a competitive experiment if the relative frequencies of different on-board clocks could be measured to a part in $10^{16}$ precision.  There is no present mission for such an experiment; but some are considering it for the future.  

Laboratory experiments which compare the gravitational and inertial forces on differently composed material samples, and the ongoing 30 years of lunar laser ranging which compares the free fall rate of Earth (1/3 iron-rich core and 2/3 silicate mantle) and differently composed Moon toward the Sun \cite {wil96} are today both confirming that gravitational free fall rates are universal to one or two parts in $10^{13}$ precision. A satellite test of UFF (the STEP experiment) is under design and development in the United States and Europe to reach a part in $10^{18}$ precision, while a quicker, but lower precision, version of such a space mission is being pursued by France (MICROSCOPE). UFF experiments are being pushed to these incredible levels of precision because no other types of experiments presently compete in their ability to detect location-dependence of fundamental physical parameters, or more generally, the presence of any new long range and extremely weak force fields in physical law which by coupling to novel attributes of matter produces violations of UFF. 

Lunar laser ranging also places the best present limits on space-time variations of Newton's gravitational parameter.  If $G$ varies cosmologically in time, the Moon's orbit will show a change in its radius which grows linear in time, and phase changes in its various oscillatory motions which grow quadratic in time.  Searching for the former effect best constrained $\dot{G}$ in the earlier years of LLR, while now the latter effects dominate.  $\dot{G}/G\: \leq\: 10^{-12}\:y^{-1}$ is LLR's present constraint. 

Theories which predict a non-zero $\dot{G}$ also generally show that $G$ will be (spatially) dependent on proximity to local matter (which produces Newtonian potential $U(\vec{r})$)
\be
G(\vec{r})\;\cong\;G_{\infty}\:\left(1+\eta\:\frac{U(\vec{r})}{c^2}\right)
\ee
The coefficient $\eta$ vanishes in general relativity but generally is non-zero in alternative metric theories of gravity.  A body such as Earth with significant internal gravitational binding energy $U_{self}$ then experiences an anomalous acceleration \cite {nor68, nor70}
\be
 \delta\vec{a}\;=\;-\;\frac{\partial lnM(G)}{\partial lnG}\:c^2\:\vec{\nabla}lnG\;\simeq\;
\left(1+\eta\:\frac{U_{self}}{c^2}\right)\;\vec{g}
\ee
LLR constrains $\eta$ to be less than $3\:10^{-4}$ which reflects the fractional precision with which general relativity's post-Newtonian structure is confirmed.
  
In order to relate location-dependence and cosmic time dependence of a physical parameter, it is considered as a function of an underlying  dynamical scalar field, $\alpha\longrightarrow\alpha(\phi)$, with this field having a lagrangian-based field equation
\be
g^{\mu\nu}\:\phi_{\mu | \nu}\;=\;4\pi G\;\frac{\partial L}{\partial\phi}\;=\;4\pi G\; \sigma(\phi,\:\vec{r},\;t)
\ee
For application to experiments looking for violation of UFF on one hand,  and to effects related to a cosmological time variation of the physical parameter on the other hand, the scalar field equation takes the respective forms
\bea
\nabla^2\:\phi\;&=&\;-4\pi G \;\sigma(\phi,\:\vec{r})/c^2 \\
\ddot{\phi}\;+3H\:\dot{\phi}\;&=&\;4\pi G\;\sigma(\phi,\:t)
\eea
with $H=\dot{R}/R$ being the Hubble expansion rate of the universe, $R(t)$ being the size parameter of the universe.  A quantitative relationship can then be sought between violations of UFF and time evolution of the physical parameter.  For two materials $i$ and $j$ the anomalous forces given by Equation (2) lead to acceleration differences
\be
\frac{|\vec{a}_i-\vec{a}_j|}{|\vec{g}_s|}\;=\;\frac{\partial ln\alpha}{\partial\phi}\;\frac{\partial ln(M_i/M_j)}{\partial ln\alpha}\;\int\sigma_s\:dV\;/\;\int\rho_s\:dV
\ee
with $\rho_s$ and $\sigma_s$ being the mass density and scalar density, respectively, of the source(s) which accelerates bodies $i$ and $j$ of respective mass-enrgies $M_i$ and $M_j$.  On the other hand, a first integration of Equation (17) yields an expression for the cosmological time evolution of $\alpha$
\bea
\frac{1}{H} \frac{\dot{\alpha}}{\alpha}\;&=&\;
\left(\frac{1}{H}\frac{\dot{\alpha}}{\alpha}\right)_{p_o}
\left(\frac{R^3H}{\partial ln\alpha/\partial\phi}\right)_{p_o}
\frac{\partial ln\alpha/\partial\phi}{R^3H} \nonumber \\
&+&\;1.5\;\frac{\partial ln\alpha}{\partial\phi}\;\int_{p_o}^p\frac{R^3(p\:')H(p\:')}{R^3H}\;\frac{\sigma(p\:')_c}{\rho(p\:')_c+(\Lambda+\dot{\phi}^2)/3\pi G}\;dp\:'
\eea
with subscript $c$ indicating averaged cosmological densities of the scalar and mass-energy sources, and $\Lambda$ being the {\it cosmological constant}.  A convenient variable $dp=dR/R=H\:dt$ has been used to parameterize cosmological time evolution, $p_o$ is a useful historic epoch and $p$ is the present (unlabeled quantities are their present values). 

As example, consider the fine structure parameter $\alpha=e^2/\hbar c$.  As previously discussed, UFF has been confirmed to about a part in $10^{13}$ among elements whose nuclear electrostatic energy contributions to mass energies typically vary by parts in $10^3$ of the whole
\be
\frac{\partial ln (M_i/M_j)}{\partial ln\alpha}\;\sim\;10^{-3}
\ee
Equation (11) can then be used to eliminate the numerically unknown factor $\partial ln\:\alpha/\partial\phi$ from the cosmological time evolution Equation (12) to produce the constraint
\be
\frac{1}{H} \frac{\dot{\alpha}}{\alpha}\;\leq\;
10^{-10}\;\int_{p_o}^p\left(\frac{R^3(p\;')H(p\:')}{R^3(p)H(p)}\right)
\;\left(\frac{\rho(p)_s}{\sigma(p)_s}\;\frac{\sigma(p\:')_c}{\rho(p\:')_c}\right)\;\frac{dp\:'}{1+(\Lambda+\dot{\phi}^2)/8\pi G\:\rho(p\:')}
\ee 
in which $p_o$ is here chosen to be a past epoch when $\dot{\alpha}$ was zero. No other direct observations have constrained $\dot{\alpha}/\alpha$ to near the level of $10^{-10}\:H$.  This raises the question of whether circumstances exist for which the cosmological time evolution observations produce interesting constraints on physical theory?  The first factor in the integrand is less than one in our expanding universe.  If the scalar density is an attribute of ordinary matter and concentrates in bodies along with ordinary matter, the second factor should also be of order one unless there is something very unusual about the present epoch.  In the ``attractor'' scenario for scalar theories \cite {danor, dapol}, it is found that the dynamical, cosmological background scalar field will evolve toward a value where it's coupling function to its source, $f(\phi)$, has a minimum (and hence zero derivative) --- if such locations exist in $f(\phi)$.  If at the present epoch this field is very close to this site, then $\sigma(p\:')/\sigma(p)$ could be a huge factor which compensates for some or much of the $10^{-10}$ factor in Equation (14).  Another way in which the integral in this equation could have a very large value is if the cosmological scalar source has little or nothing to do with ordinary matter and does not concentrate in ordinary matter sources such as stars or planets to the degree that the ordinary matter does \cite {damor}.

Cosmological time variation of a physical parameter also establishes a cosmic preferred frame and body-dependent {\it Aristotelian-like} accelerations \cite {nor90}.  If one performs a Lorentz transformation to the rest frame of a body which moves at velocity $\vec{v}$ relative to the locality's special frame in which a parameter has pure time dependence, a spatial gradient is acquired in the new frame, $\alpha(t)\longrightarrow\alpha(t'+\vec{v}\cdot\vec{r}\:'/c^2)$. From equation (2), a force then accelerates the body in proportion to its velocity in the cosmos
\be
\frac{d\vec{v}}{dt}\;\cong\;-\frac{\dot{\alpha}}{\alpha}\frac{\partial lnM}{\partial ln\alpha}\;\vec{v}
\ee
Another way to understand the origin of this acceleration rests on the time variation of the body's mass and conservation of momentum 
\be
0\;=\;\frac{d}{dt}\left(M(t)\vec{v}\right)\;=\;M\dot{\vec{v}}+\dot{M}\vec{v}\;=\;
M\dot{\vec{v}}+\frac{\dot{\alpha}}{\alpha}\frac{\partial M}{\partial ln\alpha}\vec{v}
\ee
which reproduces Equation (15).

This {\it bi-product} of a cosmically time-varying physical parameter alters the response of a binary pulsar system's orbit to a changing gravitational coupling strength $G(t)$. A near-circular binary system has period $P\;=\;2\pi L^3/(GM)^2$ plus relativistic corrections, $L$ is the angular motion quantity $|\vec{r}\times\vec{v}|$, and $M$ is the mass of the system. Under changing $G$ both $L$ and $M$ change as well, so
\be
\frac{\dot{P}}{P}\;\cong\;3\frac{\dot{L}}{L}-2\frac{\dot{G}}{G}-2\frac{\dot{M}}{M}
\ee
For gravitationally compact bodies (those having large $\partial lnM/\partial lnG$) such as neutron stars the fractional time rate of change of both $l$ and $m$ can be a few tenths of that for $G$ and have signs working in opposition to $\dot{G}$. The $\dot{P}$/$\dot{G}$ ratio is therefore generally significantly reduced for such systems. 

If the two bodies in the binary system have different gravitational compactnesses, then the system will also experience a polarizing force in the direction of the system's velocity relative to the cosmic preferred frame, and the resulting orbital perturbation can also be sought as a measure of time variation of $G$. 
  
{\bf This work was supported in part by National Aeronautics and Space Administration Contract NASW - 00011.}


\begin{thebibliography}{99}
\bibitem {ves79} Vessot R F C {\it et al} (1980) {\it Phys. Rev. Lett} {\it bf 45} 1081
\bibitem {wil96} Williams J G, Newhall X X and Dickey J O (1996) {\it Phys. Rev.} {\bf D 53} 6730
\bibitem {nor68} Nordtvedt K (1968) {\it Physical Review} {\bf 169} 1017
\bibitem {nor70} Nordtvedt K (1970) {\it Physical Review} {\it 170} 1186
\bibitem {danor} Damour T and Nordtvedt K (1993) {\it Phys. Rev. Lett} {\bf 70} 2217
\bibitem {dapol} Damour T and Polyakov A M (1994) {\it Nuclear Phys B} {\bf 423} 532
\bibitem {damor} Damour T {\it Equivalence Principle and Clocks} in Gravitational Waves and Experimental Gravity, edited by J Tran Than Van et al (World Publishers, Hanoi, 2000) 357-363
\bibitem {nor90} Nordtvedt K (1990) {\it Phys. Rev. Lett.} {\bf 65} 953
\end{thebibliography}
\end{document}